\documentclass[runningheads]{llncs}
\usepackage[T1]{fontenc}
\usepackage{graphicx,verbatim}
\usepackage{xcolor}
\usepackage{calc}
\usepackage{amsmath}
\usepackage{hyperref}

\makeatletter
\newcommand{\showfontsize}{\f@size pt}
\makeatother

\begin{document}
\title{LEAVS: An LLM-based Labeler for Abdominal CT Supervision}

\ifdefined\DOUBLEBLIND
\author{Anonymized Authors}  %% Added for anonymized MICCAI 2025 submission
\authorrunning{Anonymized Author et al.}
\institute{Anonymized Affiliations \\
    \email{email@anonymized.com}}
\else
  %% Removed for anonymized MICCAI 2025 submission
\author{Ricardo Bigolin Lanfredi\inst{1}\orcidID{0000-0001-8740-5796} \and
Yan Zhuang\inst{2,3}\orcidID{0000-0003-1756-0277} \and
Mark Finkelstein\inst{2}
\and
Praveen Thoppey Srinivasan Balamuralikrishna\inst{1}
\and
Luke Krembs\inst{4}
\and
Brandon Khoury\inst{4}\orcidID{0000-0002-9625-9353}
\and
Arthi Reddy\inst{2}\orcidID{0009-0001-7645-5613}
\and
Pritam Mukherjee\inst{1}\orcidID{0000-0002-9975-9994}
\and
Neil M. Rofsky\inst{2}
\and
Ronald M. Summers\inst{1}\orcidID{0000-0001-8081-7376}
}

\authorrunning{R. Bigolin Lanfredi et al.}
\institute{
National Institutes of Health Clinical Center, Bethesda, MD 20892, USA 
\email{rsummers@cc.nih.gov}\\
\and
Department of Diagnostic, Molecular and Interventional Radiology, Icahn School of Medicine at Mount Sinai, New York, NY 10029, USA \and
Windreich Department of Artificial Intelligence and Human Health, Icahn School of Medicine at Mount Sinai, New York, NY 10029, USA \and
Walter Reed National Military Medical
Center, Bethesda, 20892, MD, USA}
\fi

% bigolinlanfrer2@nih.gov
% **yan's email:
% mark.finkelstein@mountsinai.org
% thoppeysrinivp2@nih.gov
% Luke.d.krembs.mil@health.mil
% brandon.r.khoury.mil@health.mil 
% **arthi's email:
% pritam.mukherjee@nih.gov
% neil.rofsky@mountsinai.org
% rsummers@cc.nih.gov

% mountsinai.org; utah.edu; nih.gov; health.mil; va.gov; childrensnational.org; mssm.edu; nyulangone.org; tnmgrmu.ac.in; tn.gov.in; Uwhealth.org; uta.edu; bmc.org; mdanderson.edu;
    
\maketitle
\begin{abstract}
Extracting structured labels from radiology reports has been employed to create vision models to simultaneously detect several types of abnormalities. However, existing works focus mainly on the chest region. Few works have been investigated on abdominal radiology reports due to more complex anatomy and a wider range of pathologies in the abdomen. We propose LEAVS (Large language model Extractor for Abdominal Vision Supervision). This labeler can annotate the certainty of presence and the urgency of seven types of abnormalities for nine abdominal organs on CT radiology reports. To ensure broad coverage, we chose abnormalities that encompass most of the finding types from CT reports. Our approach employs a specialized chain-of-thought prompting strategy for a locally-run LLM using sentence extraction and multiple-choice questions in a tree-based decision system. We demonstrate that the LLM can extract several abnormality types across abdominal organs with an average F1 score of 0.89, significantly outperforming competing labelers and humans. Additionally, we show that extraction of urgency labels achieved performance comparable to human annotations. Finally, we demonstrate that the abnormality labels contain valuable information for training a single vision model that classifies several organs as normal or abnormal. We release our code and structured annotations for a public CT dataset containing over 1,000 CT volumes. 
\end{abstract}

%\textcolor{blue}{[Yan: will we release the code on Github? if so, we might need to mention it in the abstract.]}

\section{Introduction}

% However, existing works focus solely on the chest region, such as chest x-ray labelers (e.g., Smit, Akshay et al. 2020) and chest CT labelers (e.g., Hamamci, Ibrahim Ethem et al. 2024).
Several works have extracted and shared structured labels from medical reports to develop generalist vision models in radiology, with examples for chest X-rays~\cite{nihdataset} and chest CTs~\cite{sarle}. The labeling has been traditionally done by rule-based algorithms~\cite{nihdataset,chexpert,mimiccxr,sarle,similar} and supervised deep learning algorithms~\cite{chexbert}. However, recent works have employed large language models (LLMs) and shown their superiority~\cite{maplez}. Work with LLMs on extracting labels from the long and diverse CT reports has been limited to specific findings~\cite{specific1,specific2,specific3,specific4}.

\begin{figure}[t]
\includegraphics[width=\textwidth]{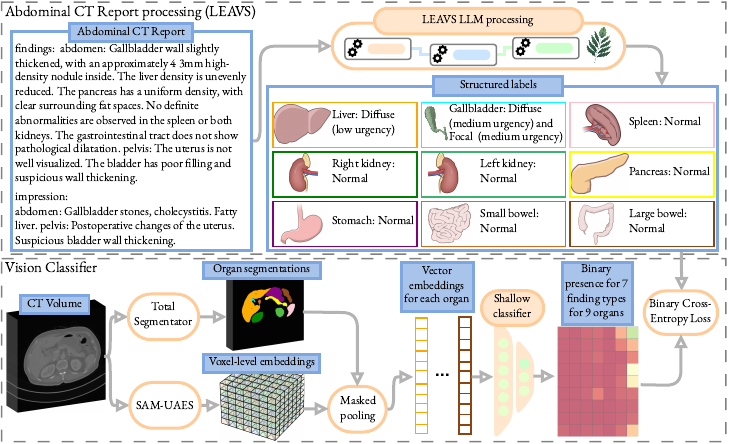}
\caption{We propose to employ LLMs to extract structured labels from CT reports to train vision models. We present an example of an input and output of our method. The report chosen for display is purposefully short. We include the pipeline we employed to classify the presence of several types of findings in several abdominal organs.} \label{overall}
\end{figure}

We propose a prompt system named LEAVS (\textbf{L}LM \textbf{E}xtractor for \textbf{A}bdominal \textbf{V}ision \textbf{S}upervision). It uses LLMs to extract comprehensive findings from abdominal CT reports for several organs and represent them as structured labels, as shown in Figure~\ref{overall}. The LEAVS prompt system is inspired by MAPLEZ (Medical report Annotations with Privacy-preserving LLM using Expeditious Zero shot answers)~\cite{maplez}, but provides several innovations when adapting it to CT reports:
\begin{itemize}
\item sentence filtration, for allowing the LLM to focus only on the parts of the long CT report that matter for its task;
\item multiple-choice questions, so the LLM picks one type for each report finding; 
\item finding type definitions for abdominal CT, chosen to cover almost all findings from reports while having enough types for separating distinct findings;
\item urgency assessment, which might be important for filtering findings and as additional information for supervision in vision models.
\end{itemize}
We demonstrate better scores than the average human for abnormality labeling and same-level scores for urgency labeling. We also employ the structured labels to train a CT vision model and demonstrate preliminary classification results, a potential step towards an approach to universal abnormality detectors for abdominal CT. Our code and annotations for the AMOS-MM dataset~\cite{amosmm} can be found at \ifdefined\DOUBLEBLIND\url{https://anonymous.4open.science/r/C3B0/}\else\url{https://github.com/rsummers11/LEAVS}\fi.

\subsection{Related works}

There have been few works extracting several abnormality types from abdominal CT reports. Islam Tushar et al.~\cite{similar} employed a rule-based algorithm to extract labels of 5 different findings per organ for three organs in the chest and abdomen and train a vision model with a private dataset. Lea Draelos et al.~\cite{sarle} employ a similar approach with the SARLE (Sentence Analysis for Radiology Label Extraction) labeler, which annotates 83 abnormalities across 52 body regions, mainly targeting chest CT scans. Both works employ rule-based algorithms, which are less flexible than LLMs since they need a new set of expertly crafted rules for every new abnormality or keyword they can identify. As we show in our work, the SARLE labeler does not perform as well in the generic task of extracting ``any abnormality'' presence in organs.

\section{Methods}
\begin{figure}[t]
\includegraphics[width=\textwidth]{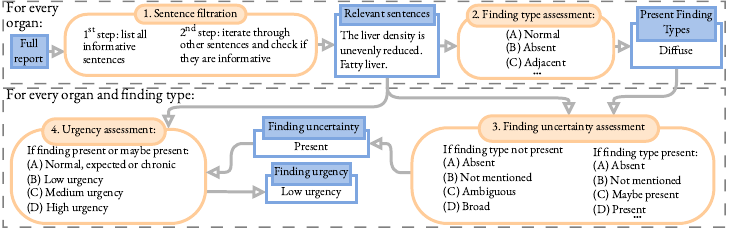}\caption{Representation of all four steps of the LEAVS LLM processing. We show data outputs for the report shown in Figure~\ref{overall} for the diffuse finding in the liver.}\label{prompt}
\end{figure}

The proposed zero-shot prompt system is presented in Figure~\ref{prompt} and contains four stages: sentence filtration, finding type assessment, finding uncertainty assessment, and urgency assessment. The first two stages are executed separately for each organ, and the remaining two for each finding type in each organ. The nine organs we evaluate are: liver, gallbladder, spleen, right kidney, left kidney, pancreas, stomach, small bowel, and large bowel. The prompt system, however, allows for the easy integration of any organ or body region.

Sentence filtration has two steps: first, the LLM is asked to list all informative sentences for an organ. Then, to increase sensitivity, we ask the LLM for each remaining sentence if they are informative for the organ. We join the informative sentences from both steps and provide only them in other prompt stages.

For the finding type assessment, the LLM is asked, in a multiple-choice question with possibly multiple answers, what types of findings from the following choices were mentioned: ``Absent'': {organ} is not present; ``Device'': {organ} has support device; ``Postsurgical'': {organ} has postsurgical changes; ``Enlarged'': {organ} is enlarged, ``Atrophy'': {organ} has atrophied, ``Anatomy'': uncommonly seen displacements, relative positionings, or shapes of the {organ}; ``Focal'': {organ} has a finding that can be measured from its borders; ``Diffuse'': {organ} has a finding without a well-defined border or shape for measurement, or that affect large regions; ``Quality'': finding about the acquisition process for the organ; ``Adjacent'': an adjacent, extrinsic finding for the {organ}; ``Normal'': {organ} is normal. ``Normal'' and ``Adjacent'' are listed to give the LLM at least one correct answer for any finding. The finding definitions are mutually exclusive, so one finding has only one type, but a single sentence or organ might have several findings. In this stage, the multiple-choice question helps the LLM to enforce the mutual exclusivity.

The LLM categorizes the uncertainty of the identified findings as: mentioned as possible, stated as positive, deemed negative or very unlikely, not directly mentioned, described with ambiguous language when comparing to a previous report of the same patient, or mentioned only for a broad anatomical area. Findings that are not present are classified among the four latter options. We classify the urgency of present or possible findings according to the definitions from Larson et al.~\cite{urgencylevels}, with an added level for non-actionable findings: normal, expected, or chronic (no action is needed); low (it could cause problems in the future); medium (it requires treatment); or high (it requires immediate treatment).

Sentences were added at the start of the prompt to ask the LLM to consider all information provided. Except for the first sentence filtration step, we employed chain-of-thought prompting (CoT) in all prompts by asking the LLM to think step-by-step and explain each sentence's medical meaning before answering. For parsing purposes, the model was asked to summarize its previous answer in a second prompt. Full prompts can be found in our code.

We tested the usefulness of the extracted labels in supervising a vision model. We trained a simple model employing a pre-trained CT embedder, SAM-UAES~\cite{samuaes}, to perform feature extraction and the segmentations of Total Segmentator~\cite{nnunet,totalsegmentator} so the model would focus on the respective organ for each output, as shown in Figure~\ref{overall}. The only trainable part of the model was the shallow classifier. The 7 finding type outputs were ``Postsurgical''+``Absent'', ``Quality'', ``Anatomy'', ``Size'': ``Enlargement''+``Atrophy'', ``Device'', ``Diffuse'', and ``Focal''.
% \begin{figure}[t]
% \includegraphics[width=\textwidth]{classifier.drawio-4.pdf}
% \caption{} \label{figclassifier}
% \end{figure}
\section{Results}
We validated LEAVS on a private dataset of 15 reports from \ifdefined\DOUBLEBLIND **************\else the NIH Clinical Center\fi, approved by the institutional IRB. We evaluated many LLMs, including the Llama 3 family~\cite{llama3}, Llama3-OpenBioLLM-70B~\cite{openbiollm}, medllama3-v20~\cite{medllama}, QwQ-32B-Preview~\cite{qwq}, Qwen2-72B-Instruct~\cite{qwen2}, and Qwen2.5-72B-Instruct~\cite{qwen25}. Our experiments employed the best validation LLM, Qwen2-72B-Instruct.
\begin{table}[t]
\centering
\caption{Scores of the proposed labeler (LEAVS), the MAPLEZ baseline~\cite{maplez} and humans. GBl: Gallbladder; RKid: Right Kidney; LKid: Left Kidney; LBow: Large Bowel; PS: postsurgical; H: Human; H Avg: average over the five humans; $N$: number of samples for score calculation (micro scores accumulate samples from non-micro rows, and humans have a reduced $N$ as each human labeled 100-150 reports and cases with disagreement between other human labelers were excluded); $N_+$: number of positive samples; F1: F1 score; 
$^{ns}$: \textit{P}$\ge$.05;
$^{\scriptscriptstyle{*}}$: \textit{P}<.05;$^{\scriptscriptstyle{*}\scriptscriptstyle{*}}$: \textit{P}<.01; $^{\mathrel{\substack{\scriptscriptstyle{\ast} \\[-1.15ex] \scriptscriptstyle{\ast\ast}}}}$: \textit{P}<.001; LEAVS$_{sub}$F1: F1 score for LEAVS in the evaluation subset of the respective human used for \textit{P} values comparisons; Pre: Precision; Rec: Recall; Spe: Specificity; MCC: Matthews correlation coefficient. We display 95\% confidence intervals for some metrics between parentheses.}\label{labeltypes}
{\fontsize{8pt}{8pt}\selectfont
\begin{tabular}{|l|l|l|r|r|r|r|r|r|r|r|}			\hline										
\multicolumn{1}{|c|}{Organ}	&	\multicolumn{1}{|c|}{Type}	&	\multicolumn{1}{|c|}{Labeler}	&	\multicolumn{1}{|c|}{$N$}	&	\multicolumn{1}{|c|}{$N_+$}	&	\multicolumn{1}{|c|}{F1}	&	\multicolumn{1}{|c|}{LEAVS$_{sub}$F1}	&	\multicolumn{1}{|c|}{Pre}	&	\multicolumn{1}{|c|}{Rec}	&	\multicolumn{1}{|c|}{Spe}	&	\multicolumn{1}{|c|}{MCC}	\\	\hline
Liver	&	Diffuse	&	LEAVS	&	200	&	31	&	.925(.844,.984)\makebox[\widthof{$^{\mathrel{\substack{\scriptscriptstyle{\ast} \\[-1.15ex] \scriptscriptstyle{\ast\ast}}}}$}][l]{$^{}$}	&	\multicolumn{1}{|c|}{-}	&	.884	&	.970	&	.976	&	.912	\\	
Liver	&	Focal	&	LEAVS	&	200	&	91	&	.963(.931,.988)\makebox[\widthof{$^{\mathrel{\substack{\scriptscriptstyle{\ast} \\[-1.15ex] \scriptscriptstyle{\ast\ast}}}}$}][l]{$^{}$}	&\multicolumn{1}{|c|}{-}	&	.929	&	1.00	&	.936	&	.932	\\	
GBl	&	Diffuse	&	LEAVS	&	200	&	36	&	.817(.700,.907)\makebox[\widthof{$^{\mathrel{\substack{\scriptscriptstyle{\ast} \\[-1.15ex] \scriptscriptstyle{\ast\ast}}}}$}][l]{$^{}$}	&\multicolumn{1}{|c|}{-}	&	.903	&	.750	&	.982	&	.788	\\	
GBl	&	Focal	&	LEAVS	&	200	&	27	&	.732(.600,.835)\makebox[\widthof{$^{\mathrel{\substack{\scriptscriptstyle{\ast} \\[-1.15ex] \scriptscriptstyle{\ast\ast}}}}$}][l]{$^{}$}	&\multicolumn{1}{|c|}{-}	&	.588	&	.966	&	.896	&	.708	\\	
GBl	&	PS	&	LEAVS	&	200	&	11	&	.957(.824,1.00)\makebox[\widthof{$^{\mathrel{\substack{\scriptscriptstyle{\ast} \\[-1.15ex] \scriptscriptstyle{\ast\ast}}}}$}][l]{$^{}$}	&\multicolumn{1}{|c|}{-}	&	1.00	&	.917	&	1.00	&	.955	\\	
Spleen	&	Size	&	LEAVS	&	200	&	21	&	.978(.919,1.00)\makebox[\widthof{$^{\mathrel{\substack{\scriptscriptstyle{\ast} \\[-1.15ex] \scriptscriptstyle{\ast\ast}}}}$}][l]{$^{}$}	&\multicolumn{1}{|c|}{-}	&	.957	&	1.00	&	.994	&	.975	\\	
RKid	&	Focal	&	LEAVS	&	200	&	62	&	.912(.850,.957)\makebox[\widthof{$^{\mathrel{\substack{\scriptscriptstyle{\ast} \\[-1.15ex] \scriptscriptstyle{\ast\ast}}}}$}][l]{$^{}$}	&\multicolumn{1}{|c|}{-}	&	.935	&	.889	&	.972	&	.874	\\	
LKid	&	Focal	&	LEAVS	&	200	&	57	&	.901(.835,.952)\makebox[\widthof{$^{\mathrel{\substack{\scriptscriptstyle{\ast} \\[-1.15ex] \scriptscriptstyle{\ast\ast}}}}$}][l]{$^{}$}	&\multicolumn{1}{|c|}{-}	&	.862	&	.948	&	.939	&	.863	\\	
LBow	&	Focal	&	LEAVS	&	200	&	38	&	.759(.640,.851)\makebox[\widthof{$^{\mathrel{\substack{\scriptscriptstyle{\ast} \\[-1.15ex] \scriptscriptstyle{\ast\ast}}}}$}][l]{$^{}$}	&\multicolumn{1}{|c|}{-}	&	.735	&	.794	&	.933	&	.702	\\	
LBow	&	PS	&	LEAVS	&	200	&	20	&	.977(.909,1.00)\makebox[\widthof{$^{\mathrel{\substack{\scriptscriptstyle{\ast} \\[-1.15ex] \scriptscriptstyle{\ast\ast}}}}$}][l]{$^{}$}	&\multicolumn{1}{|c|}{-}	&	1.00	&	.955	&	1.00	&	.974	\\	
micro	&	micro	&	LEAVS	&	2000	&	394	&	.892(.869,.913)\makebox[\widthof{$^{\mathrel{\substack{\scriptscriptstyle{\ast} \\[-1.15ex] \scriptscriptstyle{\ast\ast}}}}$}][l]{$^{}$}	&\multicolumn{1}{|c|}{-}	&	.865	&	.921	&	.965	&	.865	\\	
micro	&	micro	&	MAPLEZ	&	2000	&	394	&	.827(.799,.851)\makebox[\widthof{$^{\mathrel{\substack{\scriptscriptstyle{\ast} \\[-1.15ex] \scriptscriptstyle{\ast\ast}}}}$}][l]{$^{\mathrel{\substack{\scriptscriptstyle{\ast} \\[-1.15ex] \scriptscriptstyle{\ast\ast}}}}$}	&	.892(.869,.913)	&	.726	&	.960	&	.911	&	.789	\\	
micro	&	micro	&	H1	&	1124	&	236	&	.871(.837,.902)\makebox[\widthof{$^{\mathrel{\substack{\scriptscriptstyle{\ast} \\[-1.15ex] \scriptscriptstyle{\ast\ast}}}}$}][l]{$^{\scriptscriptstyle{*}\scriptscriptstyle{*}}$}	&	.923(.898,.946)	&	.950	&	.805	&	.989	&	.845	\\	
micro	&	micro	&	H2	&	843	&	193	&	.935(.906,.960)\makebox[\widthof{$^{\mathrel{\substack{\scriptscriptstyle{\ast} \\[-1.15ex] \scriptscriptstyle{\ast\ast}}}}$}][l]{$^{ns}$}	&	.927(.899,.952)	&	.973	&	.902	&	.992	&	.918	\\	
micro	&	micro	&	H3	&	476	&	125	&	.930(.893,.961)\makebox[\widthof{$^{\mathrel{\substack{\scriptscriptstyle{\ast} \\[-1.15ex] \scriptscriptstyle{\ast\ast}}}}$}][l]{$^{ns}$}	&	.940(.907,.968)	&	.967	&	.898	&	.989	&	.908	\\	
micro	&	micro	&	H4	&	543	&	117	&	.869(.819,.911)\makebox[\widthof{$^{\mathrel{\substack{\scriptscriptstyle{\ast} \\[-1.15ex] \scriptscriptstyle{\ast\ast}}}}$}][l]{$^{\scriptscriptstyle{*}\scriptscriptstyle{*}}$}	&	.935(.900,.964)	&	.885	&	.855	&	.970	&	.835	\\	
micro	&	micro	&	H5	&	451	&	107	&	.868(.811,.913)\makebox[\widthof{$^{\mathrel{\substack{\scriptscriptstyle{\ast} \\[-1.15ex] \scriptscriptstyle{\ast\ast}}}}$}][l]{$^{ns}$}	&	.898(.850,.936)	&	.918	&	.826	&	.977	&	.832	\\	
micro	&	micro	&	H Avg	&\multicolumn{1}{|c|}{-}	&\multicolumn{1}{|c|}{-}	&	.894(.876,.911)\makebox[\widthof{$^{\mathrel{\substack{\scriptscriptstyle{\ast} \\[-1.15ex] \scriptscriptstyle{\ast\ast}}}}$}][l]{$^{\scriptscriptstyle{*}\scriptscriptstyle{*}}$}	&	.924(.909,.938)	&	.938	&	.856	&	.983	&	.867	\\	\hline
\end{tabular}
}
\end{table}

To test our labeler, we annotated 200 reports from the validation set of AMOS-MM~\cite{amosmm}, a public dataset of abdominal CT volumes and reports with unspecified license. Humans annotated five finding types (``Quality'', ``Postsurgical''+``Absent'', ``Size'', ``Diffuse'', and ``Focal'') for nine abdominal organs. Five annotators labeled 100 to 150 reports each, with three in total per report. When evaluating LEAVS, the ground truth was the majority vote from humans for binary labels and the average of available human labels for urgency. We report results for organs/abnormalities with more than 10 positive cases in the test set. We calculated two-sided hypothesis tests of the difference in scores against the proposed labeler (LEAVS) and confidence intervals with 95\% significance employing a paired bootstrap permutation test with 2,000 samples. When evaluating humans, we only considered cases when the two other humans who labeled that specific report agreed on the presence to avoid biasing the scores. Urgency labels considered only the urgency of the other two humans. When cases were filtered for human evaluation, we compared against the LEAVS scores in the same subset for fair comparison. We use micro-F1, except when aggregating over humans. For that case and other metrics, we perform macro aggregation.
\begin{table}[t]
\centering
\caption{Evaluation of ``any abnormality'' presence labeling in each organ for proposed LEAVS labeler, MAPLEZ~\cite{maplez} and SARLE~\cite{sarle} baseline, and the average of humans. Scores for the six organs with $N_+$>10 (liver, gallbladder, spleen, kidneys, pancreas, bowels) were aggregated using micro scores. Refer to Table~\ref{labeltypes} for table symbols.}\label{labelorgans}
{\fontsize{8pt}{8pt}\selectfont
\begin{tabular}{|l|r|r|r|r|r|r|r|r|}			\hline		
\multicolumn{1}{|c|}{Labeler}	&	\multicolumn{1}{|c|}{$N$} & \multicolumn{1}{|c|}{$N_+$} & \multicolumn{1}{|c|}{F1}	&	\multicolumn{1}{|c|}{LEAVS$_{sub}$F1}	&	\multicolumn{1}{|c|}{Pre}	&	\multicolumn{1}{|c|}{Rec}	&	\multicolumn{1}{|c|}{Spe}	&	\multicolumn{1}{|c|}{MCC}	\\	\hline
LEAVS	&	1200 & 381 & .961(.946,.973)\makebox[\widthof{$^{\mathrel{\substack{\scriptscriptstyle{\ast} \\[-1.15ex] \scriptscriptstyle{\ast\ast}}}}$}][l]{$^{}$}	&\multicolumn{1}{|c|}{-}	&	.936	&	.987	&	.969	&	.942	\\	
MAPLEZ	&	1200 & 381 &	.938(.919,.954)\makebox[\widthof{$^{\mathrel{\substack{\scriptscriptstyle{\ast} \\[-1.15ex] \scriptscriptstyle{\ast\ast}}}}$}][l]{$^{\mathrel{\substack{\scriptscriptstyle{\ast} \\[-1.15ex] \scriptscriptstyle{\ast\ast}}}}$}	&	.960(.946,.974)	&	.889	&	.992	&	.942	&	.909	\\	
SARLE	&	1200 & 381 & .705(.673,.734)\makebox[\widthof{$^{\mathrel{\substack{\scriptscriptstyle{\ast} \\[-1.15ex] \scriptscriptstyle{\ast\ast}}}}$}][l]{$^{\mathrel{\substack{\scriptscriptstyle{\ast} \\[-1.15ex] \scriptscriptstyle{\ast\ast}}}}$}	&	.960(.946,.974)	&	.570	&	.921	&	.678	&	.558	\\	
H Avg	&\multicolumn{1}{|c|}{-} & \multicolumn{1}{|c|}{-} & .961(.952,.969)\makebox[\widthof{$^{\mathrel{\substack{\scriptscriptstyle{\ast} \\[-1.15ex] \scriptscriptstyle{\ast\ast}}}}$}][l]{$^{\scriptscriptstyle{*}\scriptscriptstyle{*}}$}	&	.976(.969,.983)	&	.975	&	.948	&	.984	&	.938	\\	\hline
\end{tabular}
}
\end{table}

In Table~\ref{labeltypes}, we provide the labeler evaluation. The MAPLEZ baseline employed the definitions of finding types from LEAVS. We provide Table~\ref{labelorgans} to evaluate against the rule-based labeler SARLE, which does not have the same label set. For this comparison, we evaluated seven organs (left and right kidney were joined, and small and large bowel were joined) for the presence of ``any abnormality'': ``Size''+``Focal''+``Diffuse''+``Postsurgical''+``Absent''. For SARLE, we considered only the relevant types, including the ``Other'' type, since it provided higher scores. An ablation study is presented in Table~\ref{ablation}. It also shows approximate inference time per report for each method when we employed 2$\times$A100 80 GB GPUs. We employed the vLLM library~\cite{vllm} for inference, reducing the original required time by around 90\%.
\begin{table}[t]
\centering
\caption{Ablation study for LEAVS and comparison to the employment of other LLMs. ``Finding type individually'': employing individual finding type assessment questions for each finding type instead of multiple-choice questions; ``Tree prompt (MAPLEZ)'': employing the MAPLEZ prompt ~\cite{maplez} for finding uncertainty assessment; ``Fast sentence filtration'': skipping the second step from sentence filtration; ``Min/R'' is minutes per report. All rows employed $N$=2000 and $N_+$= 394. Refer to Table~\ref{labeltypes} for table symbols.}\label{ablation}
{\fontsize{8pt}{8pt}\selectfont
\begin{tabular}{|l|r|r||l|r|r|}
\hline												
\multicolumn{1}{|c|}{Labeler}	&	\multicolumn{1}{|c|}{F1}	&	\multicolumn{1}{|c||}{Min/R}	&	\multicolumn{1}{|c|}{Labeler}	&	\multicolumn{1}{|c|}{F1}	&	\multicolumn{1}{|c|}{Min/R}	\\	\hline
LEAVS (Qwen 2 72B)	&	.892(.868,.913)\makebox[\widthof{$^{\mathrel{\substack{\scriptscriptstyle{\ast} \\[-1.15ex] \scriptscriptstyle{\ast\ast}}}}$}][l]{$^{}$}	&	17.2	&	Llama 3.3 70B &	.914(.893,.933)\makebox[\widthof{$^{\mathrel{\substack{\scriptscriptstyle{\ast} \\[-1.15ex] \scriptscriptstyle{\ast\ast}}}}$}][l]{$^{\scriptscriptstyle{*}}$}	&	20.3	\\	
No CoT	&	.879(.853,.902)\makebox[\widthof{$^{\mathrel{\substack{\scriptscriptstyle{\ast} \\[-1.15ex] \scriptscriptstyle{\ast\ast}}}}$}][l]{$^{ns}$}	&	1.4	&	Qwen 2.5 72B &	.877(.850,.899)\makebox[\widthof{$^{\mathrel{\substack{\scriptscriptstyle{\ast} \\[-1.15ex] \scriptscriptstyle{\ast\ast}}}}$}][l]{$^{ns}$}	&	21.4	\\	\cline{4-6}
Finding type individually	&	.874(.849,.898)\makebox[\widthof{$^{\mathrel{\substack{\scriptscriptstyle{\ast} \\[-1.15ex] \scriptscriptstyle{\ast\ast}}}}$}][l]{$^{\scriptscriptstyle{*}\scriptscriptstyle{*}}$}	&	16.8	&	\multicolumn{3}{c}{}	\\	
Tree prompt (MAPLEZ)	&	.879(.853,.901)\makebox[\widthof{$^{\mathrel{\substack{\scriptscriptstyle{\ast} \\[-1.15ex] \scriptscriptstyle{\ast\ast}}}}$}][l]{$^{ns}$}	&	16.1	&	\multicolumn{3}{c}{}	\\	
Fast sentence filtration	&	.893(.870,.915)\makebox[\widthof{$^{\mathrel{\substack{\scriptscriptstyle{\ast} \\[-1.15ex] \scriptscriptstyle{\ast\ast}}}}$}][l]{$^{ns}$}	&	5.8	&		\multicolumn{3}{c}{}\\	
No sentence filtration	&	.866(.841,.891)\makebox[\widthof{$^{\mathrel{\substack{\scriptscriptstyle{\ast} \\[-1.15ex] \scriptscriptstyle{\ast\ast}}}}$}][l]{$^{\scriptscriptstyle{*}}$}	&	13.0	&		\multicolumn{3}{c}{}\\	\cline{1-3}

\end{tabular}
}
\end{table}

We employ the Kendall Tau-b correlation coefficient to evaluate urgency outputs in Table~\ref{urgency}. Scores were calculated for the maximum urgency in each organ. Similarly to the AUC metric, differences in calibration do not impact these scores as long as the more urgent cases have a higher urgency than the less urgent cases. To evaluate differences in calibration, we present Table~\ref{urgencyprevalence}.
\begin{table}[t]
\centering
\caption{Scores for urgency outputs from LEAVS and humans employing the Kendall Tau-b correlation coefficient ($\tau_b$). Instead of $N_+$, we consider $N-N_{mode}$, where $N_{mode}$ is the most common ground truth urgency. Refer to Table~\ref{labeltypes} for symbols.}\label{urgency}
{\fontsize{8pt}{8pt}\selectfont
\begin{tabular}{|l|l|r|r|r|r|}
\hline		
\multicolumn{1}{|c|}{Organ}	&	\multicolumn{1}{|c|}{Labeler}	&	\multicolumn{1}{|c|}{$N$}	&	\multicolumn{1}{|c|}{$N-N_{mode}$}	&	\multicolumn{1}{|c|}{$\tau_b$}	&	\multicolumn{1}{|c|}{$LEAVS_{sub}\tau_b$}	\\	\hline
Liver	&	LEAVS	&	102	&	52	&	.612(.502,.705)\makebox[\widthof{$^{\mathrel{\substack{\scriptscriptstyle{\ast} \\[-1.15ex] \scriptscriptstyle{\ast\ast}}}}$}][l]{$^{}$}	&\multicolumn{1}{|c|}{-}	\\	
GBl	&	LEAVS	&	50	&	37	&	.635(.471,.759)\makebox[\widthof{$^{\mathrel{\substack{\scriptscriptstyle{\ast} \\[-1.15ex] \scriptscriptstyle{\ast\ast}}}}$}][l]{$^{}$}	&\multicolumn{1}{|c|}{-}	\\	
Kidneys	&	LEAVS	&	84	&	43	&	.632(.489,.734)\makebox[\widthof{$^{\mathrel{\substack{\scriptscriptstyle{\ast} \\[-1.15ex] \scriptscriptstyle{\ast\ast}}}}$}][l]{$^{}$}	&\multicolumn{1}{|c|}{-}	\\	
Bowels	&	LEAVS	&	45	&	35	&	.460(.214,.654)\makebox[\widthof{$^{\mathrel{\substack{\scriptscriptstyle{\ast} \\[-1.15ex] \scriptscriptstyle{\ast\ast}}}}$}][l]{$^{}$}	&\multicolumn{1}{|c|}{-}	\\	
Macro	&	LEAVS	&\multicolumn{1}{|c|}{-}	&\multicolumn{1}{|c|}{-}	&	.582(.503,.655)\makebox[\widthof{$^{\mathrel{\substack{\scriptscriptstyle{\ast} \\[-1.15ex] \scriptscriptstyle{\ast\ast}}}}$}][l]{$^{}$}	&\multicolumn{1}{|c|}{-}	\\	
Macro	&	H1	&\multicolumn{1}{|c|}{-}	&\multicolumn{1}{|c|}{-}	&	.606(.500,.696)\makebox[\widthof{$^{\mathrel{\substack{\scriptscriptstyle{\ast} \\[-1.15ex] \scriptscriptstyle{\ast\ast}}}}$}][l]{$^{ns}$}	&	.542(.409,.661)	\\	
Macro	&	H2	&\multicolumn{1}{|c|}{-}	&\multicolumn{1}{|c|}{-}	&	.546(.446,.633)\makebox[\widthof{$^{\mathrel{\substack{\scriptscriptstyle{\ast} \\[-1.15ex] \scriptscriptstyle{\ast\ast}}}}$}][l]{$^{ns}$}	&	.507(.393,.607)	\\	
Macro	&	H3	&\multicolumn{1}{|c|}{-}	&\multicolumn{1}{|c|}{-}	&	.717(.621,.791)\makebox[\widthof{$^{\mathrel{\substack{\scriptscriptstyle{\ast} \\[-1.15ex] \scriptscriptstyle{\ast\ast}}}}$}][l]{$^{\scriptscriptstyle{*}}$}	&	.556(.371,.682)	\\	
Macro	&	H4	&\multicolumn{1}{|c|}{-}	&\multicolumn{1}{|c|}{-}	&	.581(.466,.674)\makebox[\widthof{$^{\mathrel{\substack{\scriptscriptstyle{\ast} \\[-1.15ex] \scriptscriptstyle{\ast\ast}}}}$}][l]{$^{ns}$}	&	.492(.313,.645)	\\	
Macro	&	H5	&\multicolumn{1}{|c|}{-}	&\multicolumn{1}{|c|}{-}	&	.336(.208,.444)\makebox[\widthof{$^{\mathrel{\substack{\scriptscriptstyle{\ast} \\[-1.15ex] \scriptscriptstyle{\ast\ast}}}}$}][l]{$^{\scriptscriptstyle{*}\scriptscriptstyle{*}}$}	&	.579(.444,.692)	\\	
\multicolumn{1}{|c|}{-}	&	H Avg	&\multicolumn{1}{|c|}{-}	&\multicolumn{1}{|c|}{-}	&	.556(.505,.599)\makebox[\widthof{$^{\mathrel{\substack{\scriptscriptstyle{\ast} \\[-1.15ex] \scriptscriptstyle{\ast\ast}}}}$}][l]{$^{ns}$}	&	.533(.453,.598)	\\	\hline
\end{tabular}
}
\end{table}
\begin{table}[t]
\centering
\caption{Prevalence of each urgency output. 0: normal/chronic/expected, 1: low urgency, 2: medium urgency, 3: high urgency.  Refer to Table~\ref{labeltypes} for table symbols.}\label{urgencyprevalence}
{\fontsize{8pt}{8pt}\selectfont
\begin{tabular}{|l|r|r|r|r|}
\hline										
\multicolumn{1}{|c|}{Labeler}	&	\multicolumn{1}{|c|}{$\%_{0}$}	&	\multicolumn{1}{|c|}{$\%_{1}$}	&	\multicolumn{1}{|c|}{$\%_{2}$}	&	\multicolumn{1}{|c|}{$\%_{3}$}	\\	\hline
LEAVS	&	4.6\%	&	41.4\%	&	38.4\%	&	15.6\%	\\	
H1	&	54.3\%	&	20.8\%	&	23.9\%	&	1.0\%	\\	
H2	&	65.1\%	&	10.5\%	&	24.1\%	&	0.3\%	\\	
H3	&	35.3\%	&	36.3\%	&	27.0\%	&	1.4\%	\\	
H4	&	36.2\%	&	37.1\%	&	22.6\%	&	4.1\%	\\	
H5	&	82.4\%	&	8.2\%	&	6.3\%	&	3.1\%	\\	\hline
\end{tabular}
}
\end{table}

We trained the classifier by randomly splitting the AMOS-MM training set, which contained 1,287 reports and CT volumes, into training (80\%) and validation (20\%), labeled by LEAVS. The final hyperparameters included a learning rate of 1e-3, a batch size of 512, the AdamW~\cite{adamw} optimizer, two sequential ResNet layers~\cite{resnet} with a dropout rate of 0.9 between them, binary cross-entropy loss, and the concatenation of maximum, minimum, and average pooling. The model was validated every five epochs, and we employed the model with the best average validation AUC. The testing on the AMOS-MM validation set, labeled by the human labelers from reports, is presented in Table~\ref{classifier}. In addition to data bootstrap, we included the variation of 5 random seeds in our statistics.
\begin{table}[t]
\centering
\caption{Scores of the vision classifier trained to predict several types of abnormalities for several abdominal organs. Refer to Table~\ref{labeltypes} for table symbols.}\label{classifier}
{\fontsize{8pt}{8pt}\selectfont
\begin{tabular}{|l|l|r|r|r||l|l|r|r|r|}
\hline			
\multicolumn{1}{|c|}{Organ}	&	\multicolumn{1}{|c|}{Type}	&	\multicolumn{1}{|c|}{$N$} & \multicolumn{1}{|c|}{$N_+$}	&	\multicolumn{1}{|c||}{AUC}	&	\multicolumn{1}{|c|}{Organ}	&	\multicolumn{1}{|c|}{Type}	&	\multicolumn{1}{|c|}{$N$} & \multicolumn{1}{|c|}{$N_+$}	&	\multicolumn{1}{|c|}{AUC}	\\	\hline
Liver	&	Diffuse	& 200 &	31	&	.755(.656,.840)	&	Spleen &	Size	& 200 &	21	&	.927(.865,.968)	\\	
Liver	&	Focal	& 200 &	91	&	.678(.602,.744)	&	RKid	&	Focal	& 200 &	62	&	.602(.522,.684)	\\	
GBl	&	PS	& 200 &	11	&	.985(.953,1.00)	&	LKid	&	Focal	& 200 &	57	&	.508(.421,.591)	\\	
GBl	&	Diffuse	& 200 &	36	&	.758(.656,.850) &	LBow	&	PS	& 200 &	20	&	.689(.583,.786) \\	
GBl	&	Focal	& 200 &	27	&	.692(.576,.800)	&	LBow	&	Focal	& 200 &	38	&	.580(.473,.676)		\\	\cline{1-5}
	\multicolumn{5}{c|}{}	&	Macro	&	Macro	& \multicolumn{1}{|c|}{-} &\multicolumn{1}{|c|}{-}	&	.716(.690,.743)	\\	\cline{6-10}
\end{tabular}
}
\end{table}
%CMIMI One research fellow and two radiology residents
\section{Discussion}

%Table 1 and 2
LEAVS significantly surpasses the average human and beats two of the five human labelers in Table~\ref{labeltypes}. It tends to show higher recall than humans, missing fewer mentions but being less precise in applying the medical definitions. The LLM has higher F1 scores when compared to humans (column LEAVS$_{sub}$F1) because the subsets with which humans are being evaluated are easier since they only include cases with agreement between the other two human labelers. LEAVS surpasses the LLM baseline, MAPLEZ, increasing the F1 score by 0.065 points. 
% This gap, compared to results from Table~\ref{labelorgans}, shows that the most important LEAVS contribution is the correct classification of abnormality types. 
The MAPLEZ evaluation employs the type definitions we derived for LEAVS, which are an additional contribution not revealed in the F1 score. Furthermore, LEAVS was validated in a private dataset, a domain different from the test AMOS-MM dataset, showing the potential of the prompt system to adapt to new domains. Matthews Correlation Coefficients (MCC) were included to evaluate if the high F1 scores could be due to an imbalance of the dataset~\cite{mcc}, but they did not show large gaps compared to the F1 scores.

Table~\ref{labelorgans} evaluates the labelers in an easier and not as fine-grained task, as can be seen by the higher scores presented. The LEAVS model still significantly has the best overall performance, surpassing both baselines, SARLE and MAPLEZ. The SARLE performance is probably low because it does not cover all the possible abnormalities with its rules and label set. 

% Table 3
As shown in Table~\ref{ablation}, sentence filtration and multiple-choice questions for finding-type assessment significantly improved results, whereas using CoT and multiple-choice questions for finding uncertainty assessment led to probable, but not statistically significant, improvements. Sentence filtration probably allows the model to focus on the important parts of the long reports: the average report in the AMOS-MM training set has 16 sentences and 1,400 characters.  Even though Llama 3.3 was not the best model in validation, it was the best in the larger testing set and is a potential change to improve results. This difference might be due to domain shifts between validation and test reports. This result also shows that the method is adaptable to at least one other LLM family.

The inference time is one limitation of LEAVS since 17 minutes per report can be a limiting factor for its use in large datasets or real-time applications. Table~\ref{ablation} shows one way to speed it up: fast sentence filtration. 
% Fast sentence filtration speeds the process up compared to no sentence extraction because of the summarization of information in the assessment prompts, which reduces input prompt sizes and might reduce LLM output lengths. 
Speeding inference up with knowledge distillation~\cite{knowledge} is a future effort.

% Table 4 and 5
The results from Table~\ref{urgency} show that the urgency labeling by the LEAVS method has approximately the same quality as the labeling from the average human labeler. However, human labelers have a large variation, with a Kendall Tau-b ranging from .336 to .717. Table~\ref{urgencyprevalence} shows a considerable variation in calibration for humans, with the prevalence for the label ``normal/chronic/expected'' ranging from 35.3\% to 82.4\%. The LLM labels show an even larger deviation from the human calibration, with a prevalence of the same label of only 4.6\% and higher prevalences for higher urgencies, which might show a tendency of the LLM to be conservative in its outputs.

% Table 6
Table~\ref{classifier} shows that the vision model can learn to identify most evaluated finding types, with AUCs ranging from 0.508 to 0.985 and an average AUC of 0.716. The AUCs have potential for future improvement, but we were able to show that the extracted information is learnable. Focal finding types seem to be the hardest to learn. This difficulty might come from how the pooling is performed over the whole organ or from the relatively low resolution of the SAM-UAES output, which halves the number of voxels in each dimension. The high performance of the postsurgical finding type for the gallbladder probably happens because there was no gallbladder segmented in most of those cases, and the embedding vector was filled with 0s. Future work will investigate classification improvements from learnable embedding networks and pooling weights. We will be attentive to releases of abdominal CT datasets with reports to expand the training dataset to several domains. We also plan to explore visual attribution methods to check if models can weakly learn to localize focal findings.

\section{Conclusion}

The zero-shot use of LLMs with the LEAVS prompt system can successfully label abnormalities for several organs in abdominal CT reports, outperforming rule- and LLM-based alternatives. A supervised vision model learned some information from the structured labels, showing potential for achieving general-purpose abnormality classification in abdominal CT.

\ifdefined\DOUBLEBLIND
\else
\begin{credits}
\subsubsection{\ackname}  This work was supported by the Intramural Research Programs of the NIH Clinical Center. Y.Z is supported in part by the Eric and Wendy Schmidt AI in Human Health Fellowship Program at Icahn School of Medicine at Mount Sinai. This work utilized the computational resources of the NIH HPC Biowulf cluster. (\url{http://hpc.nih.gov})

\subsubsection{\discintname}
R.M.S.: Royalties from iCAD, ScanMed, Philips, Translation Holdings, PingAn, MGB; research support through a CRADA with PingAn. Other authors have no competing interests to declare relevant to this article's content. 

\end{credits}
\fi

\bibliographystyle{splncs04}
\bibliography{refs}

\end{document}